\documentclass[prl,showpacs,showkeys,floatfix,nofootinbib,eqsecnum,
                  preprint,12pt,tightenlines,fleqn]{revtex4}

\usepackage{amsmath,amssymb,revsymb,graphicx,dcolumn}

\begin{document}
\title{Direct Observation of a Majorana Quasiparticle Heat Capacity in  $^3$He }
\author{}

\affiliation
{}

\date{\today}
\begin{abstract}
The Majorana fermion, which acts as its own antiparticle, was predicted by Majorana in 1937 \cite{Majorana}. No fundamental particles are known to be Majorana fermions, although there are speculations that the neutrino may be one. There is also theoretical speculation that Majorana fermions may comprise a large fraction of cosmic Dark Matter. While no stable particle with Majorana properties has yet been observed, Majorana {\it quasi}particles may exist at the boundaries of topological insulators. Here we report the deviation of the time constant of a Dark-Matter bolometer which arises from the additional heat capacity of the Majorana quasiparticles.
\end{abstract}


\maketitle

The theoretical suggestion of Majorana fermions, particles whose defining property is that they are their own anti-particles, has impacted on diverse problems, ranging from neutrino physics and dark matter searches to the fractional quantum Hall effect and superconductivity. Despite this long history, the unambiguous observation of Majorana fermions remains an outstanding goal. Recent advances in the condensed-matter search for Majorana particles or quasiparticles have convinced many in the field that this quest may soon bear fruit.  Numerous proposals for Majorana equivalents have been discussed, based on various materials such as topological insulators, conventional semiconductors, ferromagnetic metals and many others. In short, there are many scenarios for experiments searching for smoking-gun Majorana signatures \cite{1,2}. In this article we report the direct observation of an unexpected increase in the superfluid $^3$He  heat capacity at the limit of extremely low temperature, which corresponds well to a Majorana heat capacity. We also propose a new cell design for an ambiguous investigation of Majorana quasiparticles in the regime where this heat capacity will dominate.

The superfluid $^3$He-B phase, one of the oldest unconventional fermionic condensates experimentally realized, is  predicted to support Majorana fermion surface states \cite{Volovik09}. Majorana fermions, characterized by the equivalence of the particle and antiparticle, have a linear dispersion relation referred to as the Majorana cone. Arising from this specific dispersion relation, the Majorana quasiparticle heat capacity will show power-law dependance on temperature, in particular with quadratic dependence on $T$ for specular scattering.  The direct observation of such a heat capacity, and in particular the dependence on temperature and magnetic field, should provide a direct verification of Majorana behaviour. Usually the heat capacity of a 2D Majorana system is very small in comparison with the heat capacity of usual quasiparticles in the bulk of superfluid $^3$He. However, with decreasing temperature the bulk heat capacity falls exponentially and at some temperature should become smaller than any Majorana contribution. To make experimental measurements of the Majorana heat capacity we propose using a Dark-Matter detector, working at the limit of ultralow temperature \cite{DMB,DMB2}. We can calculate the experimental conditions under which the Majorana heat capacity should dominate and be capable of direct measurement. With this in mind we have reexamined the results of previous Dark Matter detection experiments published in Ref. \cite{DMD2} and we find that the observed  10 $\%$ deviation of the bolometer time constant can be well explained by an additional heat capacity arising from Majorana excitations.  In other words, we can say that a Majorana heat capacity has been observed directly in this experiment!

The Dark-Matter detector was developed at the Institute Neel under the auspices of the ANR project Ultima. It consists of a box with a small calibrated hole immersed in superfluid $^3$He at extreme low-temperatures. Devices of this type were first developed at Lancaster University and are known as Black-Body radiators. Such devices have been used as a source of quasiparticle wind in the ballistic regime \cite{lanc1}. Later, they have been tested in the bolometer regime\cite{lanc2}. In the latter case, the superfluid $^3$He temperature inside the box was measured by vibrating wire thermometers. The $^3$He in the box is cooled by the emission of ballistic quasiparticles through the orifice into the surrounding $^3$He which is significantly colder. When a neutron source was placed near the refrigerator, it was shown that the each neutron, which undergoes a nuclear absorption process by a $^3$He nucleus inside the box, releases an energy of around 764 keV. Later in Grenoble this type of bolometer was redesigned to provide better energy resolution. In this case, it was found that not all the energy released by the capture process was deposited in the quasiparticles gas. A significant part of energy remained unaccounted for. This energy turned out to agree well with what would be expected for the creation of a tangle of vortices corresponding to the Kibble-Zurek scenario\cite{KZ} for Cosmic-String creation after the Big Bang \cite{Nature}. Although Cosmic Strings have not yet been observed in the Universe, the $^3$He experiments indeed demonstrate the principal elements of topological defect formation during rapid phase transitions. Later similar $^3$He detectors were developed for the direct search for Dark Matter. \cite{DMD1,DMD2,DMD3}

The main principle of the Dark Matter detector relies on the measurements of the quasiparticle density in bulk $^3$He by a micron-diameter vibrating wire. An elementary-particle event releases energy locally heating the liquid $^3$He in the bolometer container. After the event the bolometer cools and the corresponding quasiparticle density decreases. The time constant of the recovery is determined by the geometry of the box, the heat capacity and the quasiparticle velocity distribution. The sensitivity of the bolometer is calibrated by the generation of a measured pulse of quasiparticles by a second vibrating wire.  The method of measuring the detector heat capacity is described in  \cite{DMB2}.

The superfluid $^3$He heat capacity falls exponentially with temperature;

\begin{equation}
  C_{bulk} \sim V \, P_F^2\,  (\frac{\Delta}{kT})^{3/2} \exp ( - \frac{\Delta}{kT} )
   \label{Cbalk}
\end{equation}
where $P_F$ is the fermi momentum, $\Delta$ - superfluid gap $ \simeq 2kT_c$ and $V$ is the volume of the sample.

In contrast the Majorana heat capacity follows a power law;

\begin{equation}
  C_{maj} \sim A \, \xi \, P_F^2\,  (\frac{\Delta}{kT})^{-2},
   \label{Cmaj}
\end{equation}
where $\xi$ is the $^3$He-B coherence length and $A$ the surface area of the foils.

The ratio of these heat capacities, including the numerical factors, is
\begin{equation}
\frac{C_{maj}}{C_{bulk}} = \frac{\pi^3}{8\sqrt2} \frac{ \xi}{\lambda} \,   (\frac{\Delta}{kT})^{-7/2}\exp \, ( \frac{\Delta}{kT} )= F \, \frac{\xi }{\lambda} ,
   \label{Cratio}
\end{equation}

 \begin{figure}[htt]
 \includegraphics[width=0.7\textwidth]{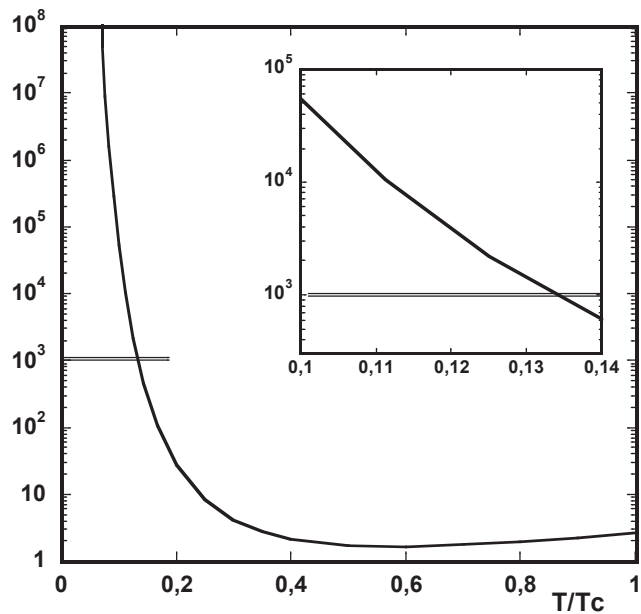}
 \caption{The ratio of the Majorana heat capacity to that of the quasiparticles $F$ from equation (0.3)(solid line) and the experimental parameter $\lambda/\xi$ for a suggested experimental cell (double line)}
 \label{profile}
\end{figure}

\begin{figure}[htt]
 \includegraphics[width=0.7\textwidth]{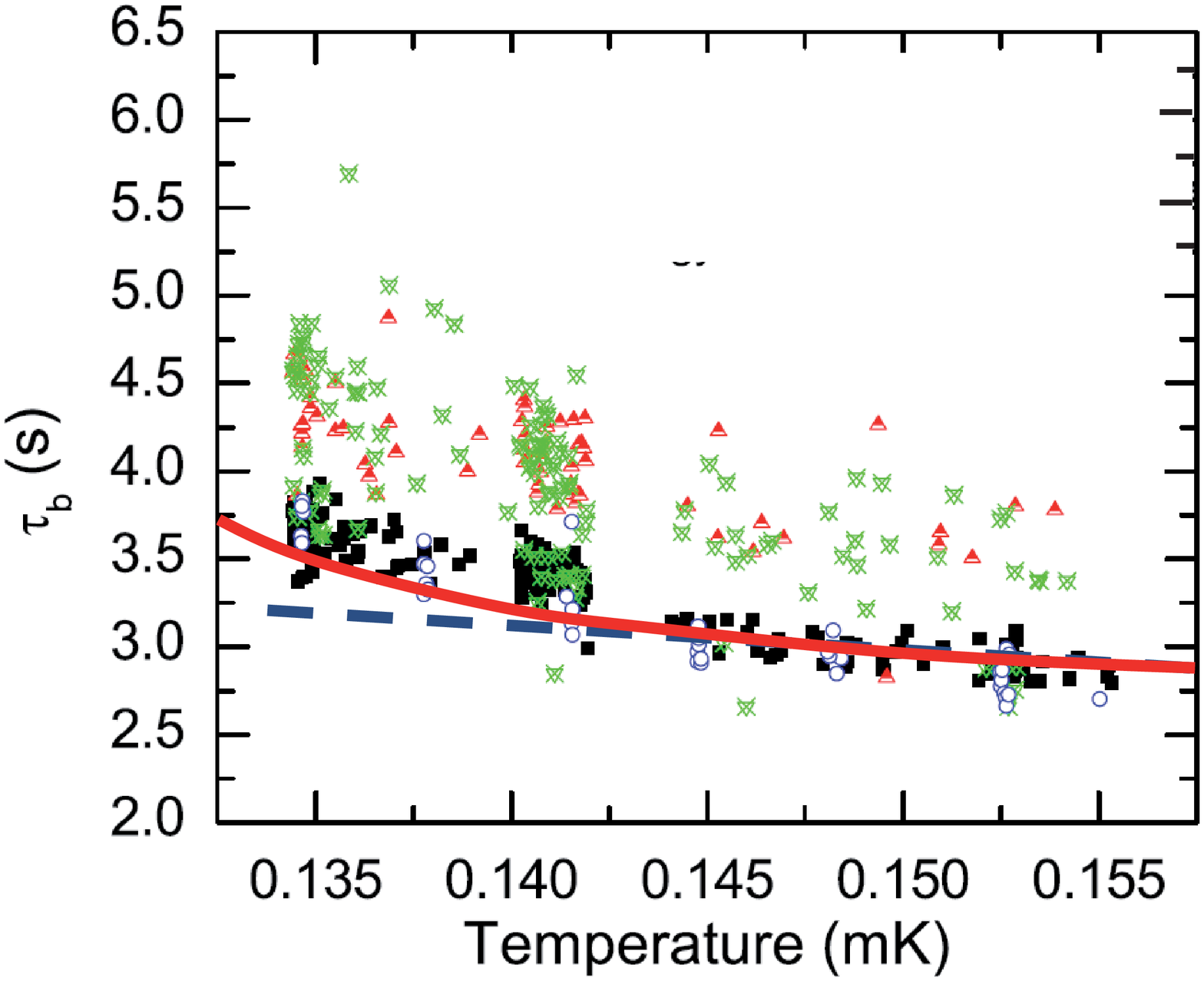}
 \caption{The time constant of the Dark Matter detector measured after heat pulses $\circ$, neutrons (squares) and muons. For details see \cite {DMD2}. The dashed line shows the theoretical dependance for the bulk quasiparticle heat capacity and solid line that with the Majorana part taken into account.  }
 \label{results}
\end{figure}
\noindent where $\lambda$ is a geometrical parameter of the experimental cell, the ratio between the volume and the surface area.

The parameter  $\lambda/\xi$ characterizes the sensitivity of the detector to Majorana quasiparticles.
In the experiments with the Dark Matter detector \cite {DMD2} the ratio $\lambda$ was about 1 mm. Consequently, the parameter $\lambda/\xi$  was about $10^4  $.  The ratio of bulk and Majorana quasiparticle heat capacities, $F$, grows very rapidly with cooling. The dependance of the quantity $F$ as a function of temperature is shown in Fig.1 by the solid line. It reaches a value of $10^4  $ at about 0.011 mK. This is the temperature, under the experimental conditions of the experiment described in \cite {DMD2} where the Majorana heat capacity should equal that of the the bulk.  Below this temperature, the heat capacity should decrease much more slowly on account of the switch from an exponential dependence to a power-law dependance. Systematic measurements of the deposited energy in the experiments \cite {DMD2} were only performed down to 0.135 mK. At this temperature, the Majorana heat capacity is estimate to be about 10 $\%$ of that of the bulk. The most sensitive method of detecting the deviation heat capacity value from that of the bulk is the measurement of the Dark-Matter detector time constant, which is the ratio of the total heat capacity to the  heat flow out of the detector  \cite {DMD2}. In this case, the effect of the fast changing quasiparticle density cancels in the two contributions, and only a square-root dependance on the temperature for the bulk remain. In Fig.2 is shown the experimental data from Fig.2 Ref.\cite {DMD2} for the time constant after various types of event. The most reliable data are for the heat pules and for neutron capture events. The muon data show a time-shifted release of energy owing to the excitation of dimers. The time constant for temperatures above 0.0145 mK corresponds well to a square-root dependance, which we have shown by a dashed line in the figure. At lower temperatures, the time constant increases significantly faster, in good agreement with the existence of a Majorana heat capacity (Eq. 3) as shown by the solid line. The good agreement of the observed additional heat capacity with that calculated for the Majorana contribution provides direct evidence of the existence of Majorana defects in superfluid $^3$He.

 \begin{figure}[htt]
 \includegraphics[width=0.5\textwidth]{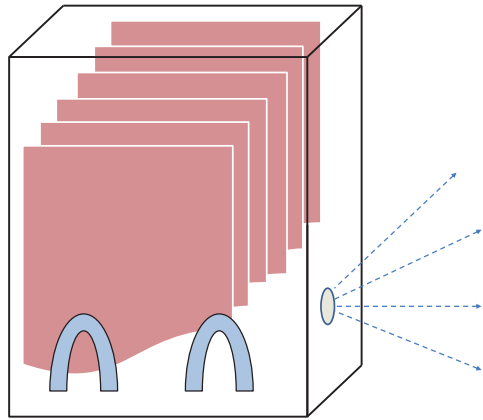}
 \caption{A schematic diagram of a Dark-Matter Detector with a stack of a foils for the direct measurement of the Majorana heat capacity.}
 \label{profile}
\end{figure}
In future we propose to place inside the Dark Matter detector a stack of foils with a separation distance of $\sim$0.1 mm as shown in Fig. 3.  In this case the  the parameter  $\lambda/\xi$ will be reduced to 10$^3$. It means that the Majorana heat capacity will equal the bulk value at  0.0135 mK as shown by the double line in the inset of Fig.1. At lower temperatures the Majorana contribution will dominate the heat capacity.  At a temperature of 100 $\mu$K the Majorana heat capacity should be be 50 times that of the bulk. This would provide the perfect conditions for the direct study of the Majorana heat capacity and its dependance on the magnetic field orientation and on the surface roughness.

In conclusion: the superfluid $^3$He may be used for to test of unusual properties of  Majorana quasiparticles. The measurements of its heat capacity is a very direct method of Magorana investigations. For today the superfluid $^3$He served as a test system for many problems of Cosmology and High energy physics \cite{DMB4,DMB5,VolovikUn}.

\vspace{2mm}

\textbf{Acknowledgements}
The authors are grateful to G.E.Volovik and G.R.Pickett for many stimulated discussions.

This work is supported  by the EU's 7th Framework Programme (FP7/2007-2013: grant agreement
\# 228464 MICROKELVIN),

\end{document}